\documentclass[12pt]{amsart}

\usepackage{ljm-auth}
\usepackage{graphicx}

\author{A.\,Kryukov, A.\,Demichev
}
\crauthor{KRYUKOV, DEMICHEV} 

\tit{Architecture of distributed data  storage for astroparticle physics}
\shorttit{Architecture of distributed data  storage} 

\begin{document}

\maketit

\address{D.V.~Skobeltsyn Institute of Nuclear Physics, M.V.~Lomonosov Moscow State University
}

\email{kryukov@theory.sinp.msu.ru, demichev@theory.sinp.msu.ru
}

\abstract{

For the successful development of the astrophysics and, accordingly, for obtaining more complete knowledge of the Universe, it is extremely important to combine and comprehensively analyze information of various types (e.g., about charged cosmic particles, gamma rays, neutrinos, etc.) obtained by using divers large-scale experimental setups located throughout the world. It is obvious that all kinds of activities must be performed continually across all stages of the data life cycle to help support effective data management, in particular, the collection and storage of data, its processing and analysis, refining the physical model, making preparations for publication, and data reprocessing taking refinement into account. In this paper we present a general approach to construction and the architecture of a system to be able to collect, store, and provide users' access to astrophysical data. We also suggest a new approach to the construction of a metadata registry based on the blockchain technology.
}

\notes{0}{
\thank{
The work presented in Section 2 of this paper was funded by the Russian Science Foundation in the framework of the grant No.~18-41-06003, and the work presented in Section 3 was funded by the Russian Science Foundation in the framework of the grant No.~18-11-00075
} 
}

\newpage

\section{Introduction}
\label{intro} 

Astroparticle physics has become a data intensive science with many terabytes of data and often with tens of measured parameters associated to each observation. While 10--15 years ago there were 1--10 Tb of data per year in astrophysics, new experimental facilities generate data sets ranging in size from 100’s to 1000’s of terabytes per year. Moreover new highly complex and massively large datasets are expected to be produced in the next decades by novel and more complex scientific instruments as well as results of data simulations needed for physical interpretation. Handling and exploring these new high volume data and making scientific research, poses a considerable technical challenge that requires the adoption of new approaches in using computing and storage resources and in organizing scientific collaborations, scientific education and science communication, where sophisticated public data centres will play the key role. These trends give rise to a number of emerging issues of big data management. An important topic for modern science in general and astroparticle physics in particular is open science, the model of free access to data (see, e.g., \cite{Dav2004}): data are accessible not solely to collaboration members but to all levels of an inquiring society, amateur or professional. This approach is especially important in the age of Big Data, when a complete analysis of the experimental data cannot be performed within one collaboration.

The work, presented in this paper, strives to develop a key component of an open science system, namely a distributed data storage (DDS) for astroparticle physics, to be able to collect, store, and analyze astrophysical data having the TAIGA~\cite{Bud2014} and KASCADE~\cite{Apel2010} experiments as examples. The novelty of the proposed approach is the development of integrated solutions: 
\begin{enumerate}
\item data combining from several astrophysical experiments in the framework of the model of open data (open science);
\item development and adaptation of distributed data storage algorithms and techniques with a common meta-catalog to provide a unified information space of the distributed repository; 
\item development and adaptation of data transmission algorithms as well as simultaneous data transmission from several data repositories thus significantly reducing load time;
\item installation of the prototype system of Big Data analysis and exporting the experimental data from KASKADE and TAIGA for testing technology of data life cycle management.
\end{enumerate}
The basic idea is the development of a Web services which will provide user access to a set of distributed data storage from a single entry point. The aggregated data from distributed sources has to be generated and transmitted to users on their requests ``on the fly'', bypassing the stage of their loading into a single data storage. Thanks to this virtualization of the storage facilities, the users see all data storages as a single one without needing to know the internal structure of each storage. Therefore, our approach avoids the construction of huge centralized data storages.  

The open data model is widely used in scientific areas. This is because, on the one hand, data acquisition is often very expensive, and on the other hand, like any experimental information, they can be reused for repeated analysis. The current growth in the volume of data received prompts us to move more eagerly to this model of the functioning of science.

One example of such a collaboration in the field of high energy physics is the open data project in European Organization for Nuclear Research (CERN; http://opendata.cern.ch). At present, only the ATLAS experiment provides open access to  $\sim 10^{14}$ proton collisions at 8 TeV. Within the framework of the project, software was developed that allows analyzing data by all interested users. Another existing project aimed at aggregating data from distributed sources is DataONE (www.dataone.org), which was originally intended for geologic and related sciences. There are three main components in the DataONE infrastructure: nodes representing data repositories (member nodes), coordinating nodes serving data management and discovery services, and the Investigator Toolkit, which contains many end-user tools for interacting with the infrastructure. Participating in the DataONE infrastructure as a node or using the Investigator Toolkit tool provides several basic services that can be used to build additional infrastructure, services, applications, and communities. Some features of our project overlap with the capabilities of the CERN and DataONE projects. But besides another application area, namely astroparticle physics, our approach differs from these projects because of data aggregation from a set of various administratively unrelated sources with the help of special adapters that allow remote storages to store data in its own formats, as well as to use different methods of data management. In our DDS environment the stored data will be accessible via overlay file systems that is very convenient for end users. In addition we suggest essentially new approach to the metadata management. In general, an analysis of related works shows that a comprehensive solution for integrating data from various sources for astroparticle physics does not currently exist.

We do not touch here attempts to build aggregation services in areas where data can be of direct commercial value. An example of such data is, in particular, satellite images for various purposes. The main obstacle for the successful operation of aggregation services in such areas is the reluctance of owners to provide data in open access precisely beause of their commercial nature. This problem is not so acute in the case of fundamental science, at least at the lapse of a certain period after data obtaining. However, such data still represent significant scientific value. One example is the recent opening of access to a significant part of its data by the project First G-APD Cherenkov Telescope (FACT, https://fact-project.org/data/). 

The proposed approach provides essential advantages over centralized solutions because of high horizontal scalability and ability for expansion of the system to the new data sources without changing its structure. As a result, this approach will allow to create reliable, economical and convenient for users and administrators system with almost unlimited possibility for increasing the volume of stored and processed astrophysical information. Though the system under development is intended for astroparticle physics, the proposed approach can be used for other scientific areas too.

Rest of the paper is organized as follows. Section~\ref{sec:2} presents basic principles of construction and DDS architecture. In Section~\ref{sec:3}, we describe the advanced approach to building a meta database for DDS. Finally, Section~\ref{sec:4} contains conclusion and future work.

\section{Basic Principles of Construction and DDS Architecture}
\label{sec:2} 

The technologies underlying the DDS system are the following: Extract-Transform-Load (ETL)~\cite{Vas2009}, Cloud technology, Web services, SaaS, REST. ETL technology includes the three main steps to proceed the data: (1)~data extraction and data verification; (2)~data transformation including data cleaning and data integration; (3)~data load and data aggregation.

The difference between the methodology proposed in this work from the usual ETL is that the selection of data from distributed sources will be generated and transmitted to users on their requests "on the fly", bypassing the stage of their loading into a single data storage. Such an approach will allow to serve users' requests for obtaining the necessary data samples, which, as a rule, require only a small part of the entire available data set for analysis. On the other hand, such a virtualization of the access to a number of storages does not require the creation of huge centralized repository. A fast data exchange is reached via caching read only filesystems CVMFS \cite{Meu} and microservice technology with REST architecture style. 

The main objects of the data model in the case of astroparticle physics are extensive air showers (EAS) recorded at experimental facilities of various types. The data model of the EAS events contains all the necessary information for subsequent physical analysis and should include: the exact time of registration of the event by the detector; time samples (histogram) of the signal from the detector in some units; service information. In addition, data on the structure of the experimental setup, for example, the coordinates of the detectors, the calibration characteristics of the detectors, other auxiliary data, should be stored as a meta-data. Indeed, to ensure that the user can actually use the data, an extensive documentation (meta-data) on how the data has been obtained is needed. Depending on the kind of data, this is at least a description of the detector and the reconstruction procedures employed. Another important aspect is the user and access management.  While there is already a basic implementation of a permission based access limitation, a useful categorization of the users into -- possibly hierarchical -- groups is needed (no administrator should manually manage privileges of single users) to effectively use it. 

Despite the fact that the facilities for registering the EAS are scattered around the world and similar to each other, they solve different physical problems. Therefore, joint processing of data will allow us to establish such properties of cosmic rays that can not be obtained from the data of individual experiments. Thus, the future model should cover as many different experiments as possible, or provide the ability to convert data using the model being developed. The XML and JSON languages is used as the description language for the data model. Further, the development of algorithms for the big data analysis of astrophysical experiments includes, among others, machine learning methods. To develop DDS prototype such programming languages as Python, JavaScript, C/C++ are used. The system itself is implemented as a set of web (micro)services using the architectural style REST. To simplify the users' work with the aggregation service, the web interfaces based on the Django framework, Node.js and JavaScriptit is implemented. The prototype of the DDS will be filled with the data of the KASCADE and TAIGA experiments for testing and system improvements. 

\begin{figure}[h]
  \includegraphics[width=0.85\textwidth]{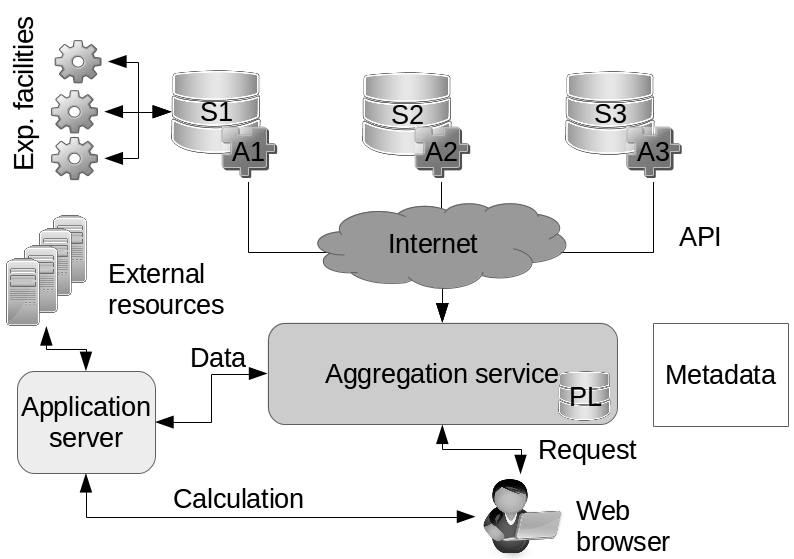}
\caption{Architecture of DDS} \label{fig:1}
\end{figure}

The architecture of DDS is presented in Figure~\ref{fig:1}.
Astrophysical data is stored in remote storages $S_n$ (n = 1,2,3, ...). These data can be of the two types: data coming from astrophysical experimental facilities (possibly after initial processing) and results of analysis of experimental data obtained after their processing by specialized application programs. Each of the remote storages can store data in its own format, as well as use different methods of data management (in particular, use different directory structure, the structure of requests for operations with files, etc.). Loading of new data into each of these real repositories is carried out using their own tools and protocols and is not within the scope of DDS. These repositories are embedded in the DDS system by means of special adapters $A_n$ that transform a specific storage API and thereby standardize the requests to repositories from the DDS side. Moreover, requests can also be of the two types: requests for operations with files and queries for searching data by their metadata. Also, the process of the embedding of a new repository must be accompanied by the loading of the relevant metadata in the metadata registry described in the next section. The box "Metadata" in Figure~\ref{fig:1} is detailed in the next section and in Figure~\ref{fig:2}.

Operations with files from the DDS side are supposed to be implemented in the spirit of the Copy-on-Write (COW) technology \cite{Dha} and overlay file system \cite{BBM,Oka}, so that the main operations are retrieving and downloading files. The ability to search data by their metadata in all storages with a single user request is a prerequisite for providing the overall common DDS environment so that for the user the system looked like a single virtual storage. A new approach to the organization of the DDS metadata registry is presented in the next section.

The central module of the system is the service of data aggregation from various data storages. The user accesses this service with a request to receive data that meet the set of criteria defined by the metadata values. The service accesses the metadata registry, which in return outputs the physical addresses of the files with data that satisfy the specified criteria. Using these addresses, the aggregation service accesses the appropriate remote repositories and downloads the required data files. In the user's request, in addition to specifying data selection criteria, an indication is given of what operations should be performed with the received files. Depending on it, the resulting set of data files is sent to one of the aggregation service submodules, which are implemented as embedded plug-ins and perform certain operations with data from the received files. The Plugin Library (PL in Figure~\ref{fig:1}) is intended for implementation of  the serialization-aggregation-deserialization process in accordance with user requirements. They must be registered in the aggregation service and run in a separate container to ensure the security of the system. In the simplest cases, the plug-ins perform a simple merging of all files into one archive, sequential processing of data in files ordered by the time of data generation, or the imposition of an additional filter, for example, on the energy of primary particles. In addition to the preinstalled plug-ins, it is possible to embed plug-ins developed by system users to solve their specific tasks. The aggregated data is then sent either to the user's local computer for further processing and analysis, or to the application server for processing on high-performance external resources.

Uploading files with processed data to remote storages is performed by the user's appropriate request to the aggregation service. This request is accompanied by the process of publishing data in the metadata repository, which consists in describing the characteristics and history of the published data (the so-called provenance metadata, see, for example, the review \cite{Zaf} and references therein). Likewise, the publishing procedure should also accompany the uploading to the storages of files with data from experimental facilities. Loading of files with experimental data is carried out under the control of managers on each storage directly or through the corresponding web service and is also accompanied by updating of the metadata register.

The performance of the system will be determined by the performance of the aggregation service. In need to increase the performance, it is possible to install several copies of this service.

\section{Metadata Registry}
\label{sec:3} 

Metadata describing data, provide context and are vital for the accurate interpretation and use of data by both humans and machines. Analysis of a problem shows a backlog in research in the field of metadata and their management systems. One of the important types of metadata is provenance (lineage, pedigree) metadata. Provenance from the point of view of computer science is a meta-information related to the history of obtaining data, starting from the source. Metadata of this type is designed to track the steps at which data were obtained, their origin, their proper storage, reproduction, for interpretation and confirmation of the scientific results obtained on their basis. Thus, provenance metadata (PMD) are important for organizing a correct research workflow to obtain reliable results.

The need for a PMD is especially essential when big data are supposed to be jointly processed by several research teams as in the case of astroparticle physics. This requires a wide and intensive exchange of data and programs for their processing and analysis, covering long periods of time, during which both the data sources and the algorithms for their processing can be modified, in particular, due to changes in sensor design, refinements in calibration, or even physical displacement. Without clear notification to all data processing participants this can give rise to catastrophic errors in the processing and analysis of data. Similar consequences can have a ``hidden'' evolution of data processing and analysis algorithms, as well as code modification, change of versions and releases of corresponding computer programs. 

Our approach is directed to the development of principles and algorithms for the formation, storage and management of the provenance metadata generated by large scientific experiments in astroparticle physics. Although a number of projects have been implemented in recent years to create systems for the support and management of metadata (see, e.g., \cite{Zaf}, \cite{Fre} and refs therein), including the provenance of data, but all the implemented solutions have security and metadata integrity issues, especially in the case of the open access model and the possibility of using metadata by organizationally unrelated or loosely coupled research communities. This is especially true for metadata of data obtained as a result of processing and analysis of primary experimental data. For brevity, we will call all such data secondary (although some of them can be obtained as a result of several processing steps). Indeed, the providers of primary data are a very limited number of experimental installations, and the corresponding provenance metadata is generated automatically. The security and integrity of the metadata database, including centralized one, for such providers can be achieved by standard methods (accounts with the appropriate rights, cryptographic keys, etc.). The situation with providers of secondary data (that is, researchers performing data processing and analysis) is significantly different. The number of such providers can be quite large and dynamically changing. Therefore, either the overhead of managing the access rights to the metadata base will be very large, or serious problems with accidental or malicious distortion of provenance metadata can occur. An example of motivation for intentional distortion of the provenance metadata may be priority considerations (for example, getting a fictitious priority in obtaining valuable results from the physical analysis of astrophysical data).

\begin{figure}[h]
  \includegraphics[width=0.85\textwidth]{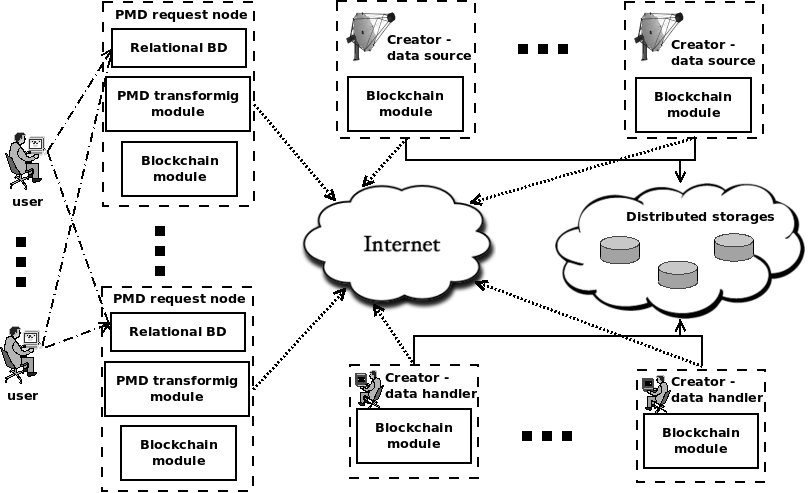}
\caption{Architecture of metadata subsystem of DDS} \label{fig:2}
\end{figure}

Fortunately, in recent years registries on the basis of blockchain technology have acquired great popularity because they have a number of important advantages (see, e.g., \cite{Ian}, \cite{Bit}) which can be successfully used in DDS. In particular,  an efficient and secure verification of the contents of large metadata structures is achieved by using in the framework of the blockchain technology the Merkle tree (hash tree) \cite{Mer} in which every leaf node is labelled with the hash of a data block and every non-leaf node is labeled with the cryptographic hash of the labels of its child nodes. The use of the Merkle cryptographic tree makes it possible to verify whether any two versions of the PMD registry are compatible: that is, a later version includes everything in the earlier version in the same order, and all new entries are received after entries in the old version. This means that no records were inserted into the registry in hindsight; no entries were changed in the registry and the registry has never been branched or bifurcated. Such a proof of consistency is important for verifying that the PMD registry was not damaged and obtained results are self-consistent too. Thus, such a distributed registry allows to monitor and restore the complete history of processing and analysis of secondary data. Figure~\ref{fig:2} shows the general architecture of the DDS metadata subsystem (the box ``Metadata'' in Figure~\ref{fig:1}). Its main feature is that data creators write the corresponding metadata into a distributed blockchain registry that provides security and data integrity, while users requesting metadata access the relational database (in read-only mode) that allows for sampling based on complex filters. Transformation of metadata from transactions of the blockchain to the relational database is carried out by the special module (PMD transforming module).

An important question is how to provide validation of the chain of blocks with transaction records in the case of PMD registry. The use of the most popular proof-of-work (PoW) method \cite{Bit} on the basis of mining is very resource-intensive, and is poorly suited for management systems for provenance metadata for the processing of scientific data. Indeed, the calculations that are performed within the framework of PoW do not serve any useful purpose, and this is a principle feature. It is very difficult to come up with a proof of work that would serve a socially useful role. Therefore, if possible, it is better to abandon it. Trying to solve these problems, a community of researchers in this field offers a variety of consensus algorithms that do not require ``work''. The choice of the algorithm heavily depends on the way of access to transaction processing. From this point of view, blockchains are classified as follows:  permissionless (public) blockchains, in which there are no restrictions on the transaction handlers (that is, accounts that can create transaction blocks); permissioned blockchains, in which transaction processing is performed by a specific list of accounts. Permissioned blockchains can form a more controlled and predictable environment than public blockchains. In contrast to the cryptocurrencies (permissionless blockchains), in the permissioned blockchains, the built-in coins are usually not used. Built-in coins are required in permissionless blockchains to provide a reward for processing transactions. The creation of blocks in an permissioned blockchains in the simplest case does not require calculations related to the work proof algorithms. In particular, the following block creation protocol, similar to the delegated stake confirmation \cite{Bit}, is possible: there is a fixed number of transaction handlers, i.e., services included in the distributed computing system (DCS); each handler owns a pair of secret and public keys, the creator of each block being determined by the mandatory digital signature of the block that is part of the block header; handlers (DCS services) create blocks in turn at fixed time intervals; the order of creation of blocks can be fixed or changed randomly after each processing cycle by all services included in the DCS; if the service for any reason can not create the block within the time interval allocated to it, it skips this cycle. In the DDS architecture, the authorized parties that create and sign the blocks are the data storages and the  metadata servers. In order to maliciously change a transaction confirmed by all the services of the DDS, the attacker must gain access to all the secret keys of the block handlers. The above protocol is theoretically even more reliable than the protocol based on the proof of work (in which case it is necessary to gain control over 51\% of the network nodes for a successful attack \cite{Fra}). It is this approach to the construction of the metadata registry that will be implemented in the DDS.

\section{Conclusion}
\label{sec:4} 

In this paper we have presented a general approach to construction and the architecture of the distributed data storage (DDS) for astroparticle physics intended for collecting, storing and distributing astrophysical data for analysis. The basic idea is the development of a Web services which will provide user access to a set of distributed data storages from a single entry point. The aggregated data from distributed sources to be generated and transmitted to users on their requests ``on the fly'', bypassing the stage of accumulating all the data into a single data storage. Thanks to this virtualization of the storage facilities, the users see all data storages as a single one. The data formats in storages entering the system can be different, the integrity being provided through the use of special adapters. In addition we have suggested a new approach to the construction of a metadata registry based on the blockchain technology. The Merkle tree will be actively used to control the data self-consistency and integrity. The new approach guarantees protection of metadata records from accidental or intentional distortions in the metadata registry. This, in turn, will significantly improve the quality and reliability of scientific results obtained on the basis of processing and analysis of big scientific data in a distributed computer environment.

To implement the DDS, a Karlsruhe-Russian initiative was put forward that united the efforts of a number of scientific institutions from Russian (SINP MSU, ISU, ISDCT SB RAS) and Germany side (KIT). The DDS will open new horizon of the computing in astroparticle physics. As a result of the developed in this work approach and architecture, a distributed system for the big astrophysical data collecting and processing will be created. A new methodology for the verification of the scientific results reliability based on the comprehensive data analysis of many types and from many sources will be developed. This also will provide an access to open astrophysical data for wide scientific community. It is worth noting that the suggested approach can be used not only in the astrophysics but also can be adapted to other scientific areas.


\end{document}